\documentclass[twocolumn,prl,superscriptaddress,showpacs]{revtex4}
\usepackage{graphicx}

\begin{document}

\title{Mechanism for the Intermittent Route to Strange Nonchaotic
Attractors}
\author{Sang-Yoon Kim}
\email{sykim@kangwon.ac.kr} \affiliation{Institute for Research in
Electronics and Applied Physics, University of Maryland, College
Park, Maryland 20742} \affiliation{Department of Physics, Kangwon
National University, Chunchon, Kangwon-Do 200-701, Korea}
\author{Woochang Lim}
\affiliation{Department of Physics, Kangwon National University,
Chunchon, Kangwon-Do 200-701, Korea}
\author{Edward Ott}
\thanks{Department of Physics and Department of Electrical and
Computing Engineering, University of Maryland, College Park,
Maryland 20742} \affiliation{Institute for Research in Electronics
and Applied Physics, University of Maryland, College Park,
Maryland 20742}

\begin{abstract}
Intermittent strange nonchaotic attractors (SNAs) appear typically
in quasiperiodically forced period-doubling systems. As a
representative model, we consider the quasiperiodically forced
logistic map and investigate the mechanism for the intermittent
route to SNAs using rational approximations to the quasiperiodic
forcing. It is thus found that a smooth torus is transformed into
an intermittent SNA via a phase-dependent saddle-node bifurcation
when it collides with a new type of ``ring-shaped'' unstable set.
Besides this intermittent transition, other transitions such as
the interior, boundary, and band-merging crises may also occur
through collision with the ring-shaped unstable sets. Hence the
ring-shaped unstable sets play a central role for such dynamical
transitions. Furthermore, these kinds of dynamical transitions
seem to be ``universal,'' in the sense that they occur typically
in a large class of quasiperiodically forced period-doubling
systems.
\end{abstract}

\maketitle

Recently much attention has been paid to the study of
quasiperiodically forced systems because of the generic appearance
of strange nonchaotic attractors (SNAs) \cite{PNR}. SNAs were
first described by Grebogi et al. \cite{Greb} and since then have
been extensively investigated both numerically
\cite{MD,Ka,HH,PD,PD2,Kuz,CGS,NK,YL,PMR,Witt,VL,NPR,OF} and
experimentally \cite{Exp}. SNAs exhibit some properties of regular
as well as chaoic attractors. Like regular attractors, their
dynamics is nonchaotic; like typical chaotic attractors, they
exhibit fractal phase space structure. Furthermore, SNAs are
related to Anderson localization in the Schr$\ddot{\rm o}$dinger
equation with a spatially quasiperiodic potential \cite{QS}, and
they may have a practical application in secure communication
\cite{SC}. Therefore, dynamical transitions in quasiperiodically
forced systems have become a topic of considerable current
interest. However, their mechanisms are still much less understood
than those in unforced or periodically forced systems.

Here we are interested in the dynamical transition to SNAs
accompanied by intermittent behavior, as reported in \cite{PMR}.
As a parameter passes a threshold value, a smooth torus is
abruptly transformed into an intermittent SNA. Near the transition
point, the intermittent dynamics on the SNA was characterized in
terms of the average interburst time and the Lyapunov exponent.
This route to an intermittent SNA is quite general and has been
observed in a number of quasiperiodically forced period-doubling
maps and flows (e.g., see \cite{Witt,VL}). It has been suggested
\cite{NPR} that the observed intermittent behavior results through
interaction with an unstable orbit. However, in the previous work,
such an unstable orbit was not located, and thus the bifurcation
mechanism for the intermittent transition still remains unclear.

In this paper, we study the underlying mechanism of the
intermittency in the quasiperiodically forced logistic map $M$
\cite{HH} which is a representative model for the
quasiperiodically forced period-doubling systems:
\begin{equation}
M: \left \{
\begin{array}{l}
x_{n+1} = (a + \varepsilon \cos 2 \pi \theta_{n}) x_{n}
(1-x_{n}), \\
\theta_{n+1} = \theta_{n} + \omega \;\;(\rm{mod}\;1),
\end{array}
\right.  \label{eq:QLM}
\end{equation}
where $x \in [0,1]$, $\theta \in S^1$, $a$ is the nonlinearity
parameter of the logistic map, and $\omega$ and $\varepsilon$
represent the frequency and amplitude of the quasiperiodic
forcing, respectively. We set the frequency to be the reciprocal
of the golden mean, $\omega = (\sqrt{5}-1)/2$. The intermittent
transition is then investigated using the rational approximations
(RAs) to the quasiperiodic forcing. For the case of the inverse
golden mean, its rational approximants are given by the ratios of
the Fibonacci numbers, $\omega_k = F_{k-1} / F_k$, where the
sequence of $\{ F_k \}$ satisfies $F_{k+1} = F_k + F_{k-1}$ with
$F_0 = 0$ and $F_1 = 1$. Instead of the quasiperiodically forced
system, we study an infinite sequence of periodically forced
systems with rational driving frequencies $\omega_k$. We suppose
that the properties of the original system $M$ may be obtained by
taking the quasiperiodic limit $k \rightarrow \infty$. Using this
technique we observe a new type of invariant unstable set, which
will be referred to as the ``ring-shaped'' unstable set in
accordance with its geometry. When a smooth torus (corresponding
to an ordinary quasiperiodic attractor) collides with this
ring-shaped unstable set, a transition to an intermittent SNA is
found to occur.

We also note that the quasiperiodically forced logistic map $M$ is
noninvertible, because its Jacobian determinant becomes zero along
the critical curve, $L_0 = \{ x=0.5,\; \theta \in [0,1) \}$.
Critical curves of rank $k$, $L_k$ $(k=1,2,\dots)$, are then given
by the images of $L_0$, [i.e., $L_k = M^k (L_0)$]. Segments of
these critical curves can be used to define a bounded trapping
region of the phase space, called an ``absorbing area,'' inside
which, upon entering, trajectories are henceforth confined
\cite{Mira}. It is found that the newly-born intermittent SNA
fills the absorbing area. Hence the global structure of the SNA is
determined by the critical curves.

\begin{figure}
\includegraphics[width=\columnwidth]{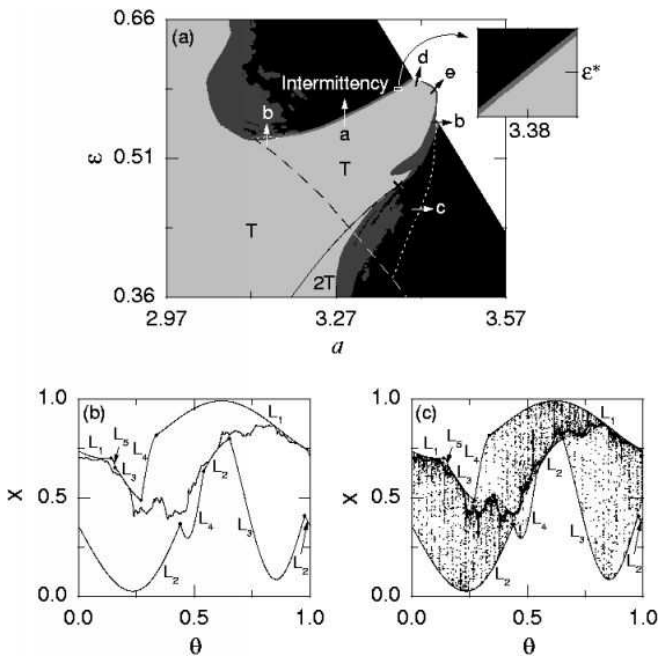}
\caption{(a) Phase Diagram in the $a-\varepsilon$ plane. Regular,
chaotic, SNA, and divergence regimes are shown in light gray,
black, gray (or dark gray), and white, respectively. To show the
region of existence (gray) of the intermittent SNA occurring
between $T$ (light gray) and the chaotic attractor region (black),
a small box near $(a,\varepsilon) = [3.38,\varepsilon^*
(=0.584\,726\,781)]$ is magnified. Through interaction with the
ring-shaped unstable set born when passing the dashed line,
typical dynamical transitions such as the intermittency (route
$a$) and the interior (routes $b$ and $c$; the dotted line is an
interior crisis line) and boundary (routes $d$ and $e$) crises may
occur. Here the torus and the doubled torus are denoted by $T$ and
$2T$, and the solid line represents a torus doubling bifurcation
curve. (b) Smooth torus inside an absorbing area with boundary
formed by segments of the critical curves $L_k$ $(k=1,\dots,5)$
for $a=3.38$ and $\varepsilon=0.584\, 7$. (c) SNA filling the
absorbing area for $a=3.38$ and $\varepsilon=0.584\,75$. For other
details, see the text. \label{fig:PD}}
\end{figure}
Figure \ref{fig:PD}(a) shows a phase diagram in the
$a-\varepsilon$ plane. Each phase is characterized by the Lyapunov
exponent $\sigma_x$ in the $x$-direction as well as the phase
sensitivity exponent $\delta$. The exponent $\delta$ measures the
sensitivity with respect to the phase of the quasiperiodic forcing
and was introduced in \cite{PD} to characterize the strangeness of
an attractor. A smooth torus that has a negative Lyapunov exponent
without phase sensitivity $(\delta=0)$ exists in the region
denoted by $T$ and shown in light gray. Upon crossing the solid
line, the smooth torus becomes unstable and bifurcates to a smooth
doubled torus in the region denoted by $2T$. Chaotic attractors
with positive Lyapunov exponents exist in the region shown in
black. Between these regular and chaotic regions, SNAs that have
negative Lyapunov exponents with high phase sensitivity
$(\delta>0)$ exist in the regions shown in gray and dark gray.
Consistent with their positive phase sensitivity exponent
$\delta$, these SNAs are observed to have fractal structure
\cite{PD}. Here we restrict our considerations only to the
intermittent SNAs that exist in the thin gray region [e.g., see a
magnified part in Fig.~\ref{fig:PD}(a)]. (In the dark-gray region,
nonintermittent SNAs, born through other mechanisms, such as
gradual fractalization \cite{NK} and torus collision \cite{HH},
exist.) Note that this phase diagram is typical for
quasiperiodically forced period-doubling systems
\cite{PMR,Witt,VL,NPR,OF,Kim}, and its main interesting feature is
the existence of the ``tongue'' of quasiperiodic motion that
penetrates into the chaotic region and separates it into upper and
lower parts. We also note that this tongue lies near the terminal
point (denoted by the cross) of the torus doubling bifurcation
curve. When crossing the upper boundary of the tongue, a smooth
torus is transformed into an intermittent SNA that exists in the
thin gray region. Hereafter this intermittent route to SNAs will
be referred to as the route $a$ [see Fig.~\ref{fig:PD}(a)].

As an example, we consider the case $a=3.38$. Figure
\ref{fig:PD}(b) shows a smooth torus with $\sigma_x = -0.059$ for
$\varepsilon=0.584\,7$ inside an absorbing area whose boundary is
formed by segments of the critical curves $L_k$ $(k=1, \dots, 5)$
(the dots indicate where these segments connect). We also note
that the smooth unstable torus $x=0$ and its first preimage $x=1$
form the basin boundary of the smooth torus in the $\theta - x$
plane. However, as $\varepsilon$ passes a threshold value
$\varepsilon^*$ $(=0.584\,726\,781)$, a transition to an
intermittent SNA occurs. As shown in Fig.~\ref{fig:PD}(c) for
$\varepsilon=0.584\,75$, the newly-born intermittent SNA with
$\sigma_x = -0.012$ and $\delta = 19.5$ appears to fill the
absorbing area, and its typical trajectory spends most of its time
near the former torus with sporadic large bursts away from it.
This intermittent transition may be expected to have occurred
through collision of the smooth attracting torus with an unstable
orbit. However, the smooth unstable torus $x=0$ cannot interact
with the smooth stable torus, because it lies outside the
absorbing area. Hence we search inside the absorbing area for an
unstable orbit that might collide with the smooth stable torus.

\begin{figure}
\includegraphics[width=\columnwidth]{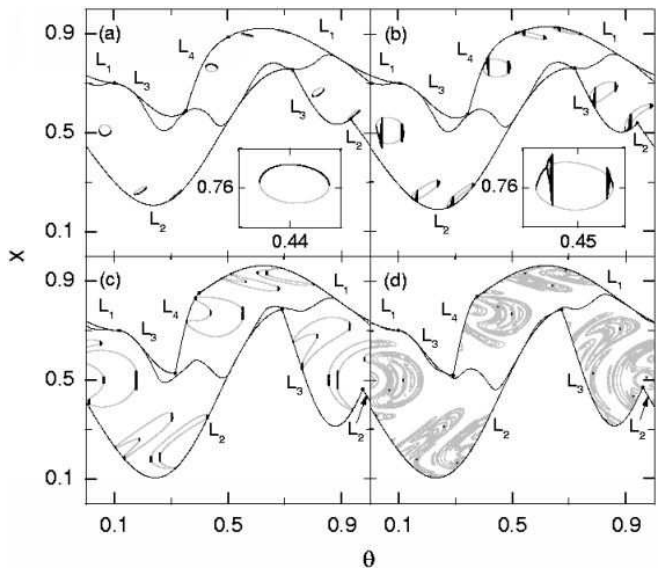}
\caption{Ring-shaped unstable sets in the RA of level $6$ for (a)
$a=3.246$ and $\varepsilon=0.446$, (b) $a=3.26$ and
$\varepsilon=0.46$, and (c) $a=3.326$ and $\varepsilon=0.526$. (d)
Ring-shaped unstable set in the RA of level $8$ for $a=3.326$ and
$\varepsilon=0.526$. These ring-shaped unstable sets exist inside
the absorbing area with boundary formed by portions of the
critical curves $L_k$ $(k=1,\dots,4)$ \label{fig:RUS}}
\end{figure}

Using RAs we find a new type of ring-shaped unstable set that
causes the intermittent transition through collision with the
smooth torus. When passing the dashed curve in
Fig.~\ref{fig:PD}(a), such a ring-shaped unstable set appears via
a phase-dependent saddle-node bifurcation. This bifurcation has no
counterpart in the unforced case. (The dashed line is numerically
obtained for the sufficiently large level $k=10$ of the RAs.) For
each RA of level $k$, a periodically forced logistic map with the
rational driving frequency $\omega_k$ has a periodic or chaotic
attractor that depends on the initial phase $\theta_0$ of the
external force. Then the union of all attractors for different
$\theta_0$ gives the $k$th approximation to the attractor in the
quasiperiodically forced system. As an example, consider the RA of
low level $k=6$. As shown in Fig.~\ref{fig:RUS}(a) for $a=3.246$
and $\varepsilon=0.446$, the RA to the smooth torus, consisting of
stable orbits with period $F_6$ $(=8)$, exists inside an absorbing
area bounded by segments of the critical curves $L_k$
$(k=1,\dots,4)$.
Note also that a ring-shaped unstable set, born via a
phase-dependent saddle-node bifurcation and composed of 8 small
rings, lies inside the absorbing area. At first, each ring
consists of the stable (shown in black) and unstable (shown in
gray) orbits with the forcing period $F_6$ $(=8)$ [see the inset
in Fig.~\ref{fig:RUS}(a)]. However, as the parameters increase, a
chaotic attractor appears through period doublings of the stable
periodic orbit, and then it disappears through collision with the
unstable $F_6$-periodic orbit (see Fig.~\ref{fig:RUS}(b) for
$a=3.26$ and $\varepsilon=0.46$). With further increase in the
parameters, both the size and shape of the rings change, and for
sufficiently large parameters, each ring consists of a large
unstable part (shown in gray) and a small attracting part (shown
in black), as shown in Fig.~\ref{fig:RUS}(c) for $a=3.326$ and
$\varepsilon=0.526$. [Note that the unstable part (plotted in the
gray) of each ring consists of only unstable orbits whose period
is the same as that of the forcing, i.e., $F_k$.] For the same
parameter values as in Fig.~\ref{fig:RUS}(c), we increase the
level of the RA to $k=8$. Then the number of rings $(=336)$
increases significantly, and the unstable part [shown in gray and
consisting of unstable orbits with period $F_8$ $(=21)$] becomes
dominant, because the attached attracting part (shown in black)
becomes negligibly small [see Fig.~\ref{fig:RUS}(d)]. In this way,
with increasing the level $k$ the ring-shaped unstable set
consists of a larger number of rings with smaller attracting part.
Hence, in the quasiperiodic limit these ring-shaped unstable sets
seem to form a complicated unstable set composed of only unstable
orbits.

In what follows we use RAs to explain the mechanism for the
intermittent transition occurring in Figs.~\ref{fig:PD}(b) and
\ref{fig:PD}(c) for $a=3.38$.
\begin{figure}
\includegraphics[width=\columnwidth]{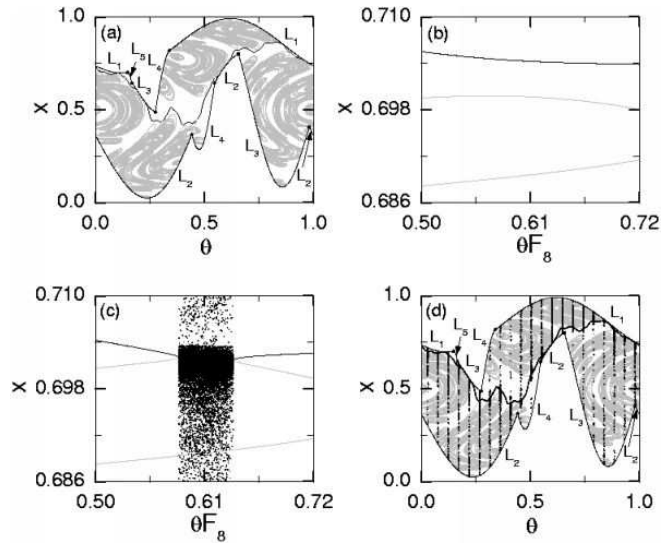}
\caption{(a) and (b) RA to the smooth torus (shown in black) and
the ring-shaped unstable set (shown in gray) in the RA of level
$8$ $(F_8 = 21)$ for $a=3.38$ and $\varepsilon=0.586$. (c) and (d)
8th RA to the SNA for $a=3.38$ and $\varepsilon=0.586\,4$. For
more details, see the text. \label{fig:SNA}}
\end{figure}
Figures \ref{fig:SNA}(a) and \ref{fig:SNA}(b) show that, inside
the absorbing area, the ring-shaped unstable set (shown in gray)
lies very close to the smooth torus (shown in black) for
$\varepsilon=0.586$ in the RA of level $k=8$. As $\varepsilon$
passes a threshold value $\varepsilon_8$ $(=0.586\, 366)$, a
phase-dependent saddle-node bifurcation occurs via a collision
between the smooth torus and the ring-shaped unstable set, and
then ``gaps,'' where no orbits with period $F_8$ $(=21)$ exist,
are formed. A magnified gap is shown in Fig.~\ref{fig:SNA}(c) for
$\varepsilon = 0.584\,6$. Note that this gap is filled by
intermittent chaotic attractors together with orbits with period
higher than $F_8$ embedded in very small windows. As shown in
Fig.~\ref{fig:SNA}(d), the RA to the whole attractor consists of
the union of the periodic component and the intermittent chaotic
component, where the latter occupies the $21$ gaps in $\theta$ and
is vertically bounded by segments of the critical curves $L_k$
$(k=1,\dots,5)$. However, the periodic component is dominant: the
average Lyapunov exponent $(<\sigma_x> = -0.09)$ is negative,
where $<\cdots>$ denotes the average over the whole $\theta$. We
note that Fig.~\ref{fig:SNA}(d) seems to be similar to
Fig.~\ref{fig:PD}(c), although the level $k=8$ is low. Increasing
the level up to $k=15$ we find that the threshold value
$\varepsilon_k$ at which the phase-dependent saddle-node
bifurcation occurs converges to the quasiperiodic limit
$\varepsilon^*$ $(=0.584\,726\,781)$ in an algebraic way, $|\Delta
\varepsilon_k| \sim F_k^{-\alpha}$, where $\Delta \varepsilon_k =
\varepsilon_k - \varepsilon^*$ and $\alpha \simeq 2.2$ \cite{Kim}.
Furthermore, in the quasiperiodic limit $k \rightarrow \infty$ the
RA to the attractor has a dense set of gaps that are filled by
intermittent chaotic attractors and bounded by portions of the
critical curves. Thus, an intermittent SNA, containing the
ring-shaped unstable set and filling the absorbing area, appears,
as shown in Fig.~\ref{fig:PD}(c). In addition, we find that as
$\varepsilon$ passes another threshold value $\varepsilon_c$
$(=0.584\,8)$, this SNA is transformed into a chaotic attractor
with a positive Lyapunov exponent. Using the RA, this transition
to chaos may also be explained. For each RA, its angle averaged
Lyapunov exponent $<\sigma_x>$ is given by the sum of the
``weighted'' Lyapunov exponents of the periodic and chaotic
components of the RA, $\Lambda_p$ and $\Lambda_c$, (i.e.
$<\sigma_x> = \Lambda_p + \Lambda_c)$, where $\Lambda_{p(c)}
\equiv M_{p(c)}\, <\sigma_x>_{p(c)}$, and $M_{p(c)}$ and
$<\sigma>_{p(c)}$ are the Lebesgue measure in $\theta$ and average
Lyapunov exponent of the periodic (chaotic) component,
respectively. After passing a threshold value where the magnitude
of $\Lambda_p$ and $\Lambda_c$ are balanced, the chaotic component
becomes dominant, and hence a chaotic attractor appears. (More
details will be given in Ref.~\cite{Kim}.)

To sum up, using RAs we have found the mechanism for the
intermittent route to SNAs in the quasiperiodically forced
logistic map. When a smooth torus makes a collision with a new
type of ring-shaped unstable set, a transition to an intermittent
SNA, bounded by segments of critical curves, occurs via a
phase-dependent saddle-node bifurcation. Other typical transitions
such as interior [routes b and c in Fig.~\ref{fig:PD}(a)] and
boundary [routes d and e in Fig.~\ref{fig:PD}(a)] crises may also
occur near the main tongue through interaction with the
ring-shaped unstable set \cite{Kim}. Furthermore, as $\varepsilon$
decreases toward zero, similar tongues appear successively near
the terminal points of the higher-order torus-doubling
bifurcations \cite{Kaneko}, and band-merging crises may also occur
through collision with the ring-shaped unstable set \cite{Kim}.
Consequently ring-shaped unstable sets play a central role for
dynamical transitions occurring near the tongues. Finally, we note
that these kinds of dynamical transitions seem to be
``universal,'' in that we observe that they occur in typical
quasiperiodically forced period-doubling systems of different
nature, such as the quasiperiodically forced H\'{e}non map, ring
map, and pendulum \cite{Kim}.

\begin{acknowledgments}
S.Y.K. thanks A. Jalnine for fruitful discussions on dynamical
transitions in quasiperiodically forced systems. This work was
supported by the Korea Research Foundation (Grant No.
KRF-2001-013-D00014) and the National Science Foundation.
\end{acknowledgments}

\end{document}